\documentclass[aps,prd,twocolumn,superscriptaddress,10pt]{revtex4-2}

\usepackage[LGR,T1]{fontenc}

\usepackage{amsmath,amsfonts}
\usepackage{graphicx}
\usepackage{bbm}
\usepackage{amssymb}
\usepackage{mathrsfs}
\usepackage{slashed}
\usepackage[linktocpage=true,
colorlinks=true,
urlcolor=blue,
anchorcolor=blue,
citecolor=blue,
filecolor=blue,
linkcolor=blue,
menucolor=blue
]{hyperref}
\usepackage[normalem]{ulem}
\usepackage{soul}
\usepackage[dvipsnames]{xcolor}
\usepackage{textcomp}    
\usepackage{textgreek}
\usepackage{tikz}
  \usepackage[justification=justified]{caption}
  \usetikzlibrary {arrows.meta}
\newcommand{\cf}{\text{\fontfamily{kurier}\selectfont\texteta}}
\newcommand{\kv}{\text{\fontfamily{kurier}\selectfont\textxi}}

\newcommand{\ex}{\mathrm{e}}

\newcommand{\dd}{\mathrm{d}}
\newcommand{\ii}{\mathrm{i}}

\newcommand{\vol}{\mathrm{vol}}

\newcommand\cC{\mathcal{C}}
\newcommand\cD{\mathcal{D}}
\newcommand\cF{\mathcal{F}}

\newcommand\cL{\mathcal{L}}
\newcommand\cM{\mathcal{M}}
\newcommand\cN{\mathcal{N}}
\newcommand\cO{\mathcal{O}}

\newcommand{\vBH}{\mathtt{v}}
\newcommand{\ls}{\ell_s}

\newcommand{\hook}{\mathbin{\rule[.2ex]{.4em}{.03em}\rule[.2ex]{.03em}{.9ex}}}

\begin{document}

\title{Probing black holes with equivariant localization}

\author{Pietro Benetti Genolini}
\affiliation{D\'epartement de Physique Th\'eorique, Universit\'e de Gen\`eve, 24 quai Ernest-Ansermet, 1211 Gen\`eve, Suisse}

\author{Christopher Couzens}
\author{Alice L\"uscher}
\affiliation{Mathematical Institute, University of Oxford, Woodstock Road, Oxford, OX2 6GG, United Kingdom}


\begin{abstract}
\noindent We introduce equivariant localization as a method for computing the action of probe branes in supergravity backgrounds. We apply it to supersymmetric probe D3-branes in type IIB supersymmetric spacetimes obtained by uplifting the Kerr--Newman-AdS$_5$ black hole on a toric Sasaki--Einstein space. Depending on the cycles they wrap, such branes represent non-perturbative corrections to or defect operator insertions in the superconformal index of a large family of four-dimensional $\mathcal{N}=1$ quiver superconformal field theories. The resulting action reduces to equivariant integrals and can be evaluated entirely from toric data.
\end{abstract}

\maketitle


\section{Introduction\label{sec:intro}}

Supersymmetric black holes in AdS$_5$ provide a remarkable arena in which to test holography beyond the semiclassical regime. While the five-dimensional geometry encodes the black holes themselves, uplifting to ten-dimensional IIB supergravity allows us to introduce a rich variety of extended supersymmetric objects, corresponding to D3-branes wrapping supersymmetry-preserving cycles. These branes probe aspects of the microscopic structure that are invisible from the purely five-dimensional perspective. Depending on the geometry they wrap, they represent different physical objects, including non-perturbative contributions to the black hole index and holographic duals to supersymmetric defects in the dual field theory. In all such cases, their contribution is governed by an integral over a calibrated cycle in the ten-dimensional geometry. Evaluating these integrals is a physically meaningful problem, but also quite non-trivial: the relevant cycle is embedded in the fibration of the internal space over the AdS black hole, and a direct evaluation quickly becomes unwieldy. The aim of this note is to show that equivariant localization provides a natural framework for carrying out this computation.

Equivariant localization techniques have emerged as a new method for computing observables in supergravity without having to solve the equations of motion \cite{BenettiGenolini:2023kxp}. The underlying idea is that supersymmetric solutions come equipped with a symmetry, while physically interesting observables are equivariant integrals with respect to it. These integrals localize, receiving contributions only from the fixed loci of the symmetry. In the context of supersymmetric supergravity solutions with an asymptotically AdS factor and an internal space, applications have mostly focused on integrals over one of these factors separately, as in the computation of on-shell actions in supergravity \cite{BenettiGenolini:2023kxp, BenettiGenolini:2024kyy, Suh:2024asy, BenettiGenolini:2024xeo,  BenettiGenolini:2024hyd, Crisafio:2024fyc, Colombo:2025ihp, Gautason:2025per, Couzens:2025ghx, Cassani:2025iix, BenettiGenolini:2025icr, Colombo:2025yqy, Park:2025fon, BenettiGenolini:2026qdm, Gaar:2026nqq}, or of extremization functionals (such as trial $c$, $F$, and $a$ functions)
\cite{BenettiGenolini:2023kxp, Martelli:2023oqk, BenettiGenolini:2023yfe, BenettiGenolini:2023ndb,  Colombo:2023fhu, Couzens:2024vbn, Cassia:2025jkr, Couzens:2025nxw,Couzens:2026qne}. Here, we show that equivariant localization can also be used to evaluate integrals over calibrated cycles extending in the full ten-dimensional geometry.

We consider supersymmetric solutions of five-dimensional minimal gauged supergravity uplifted to type IIB supergravity on a toric Sasaki--Einstein five-manifold \cite{Buchel:2006gb}, which admits an effective U$(1)^3$ action preserving the Sasakian structure. We introduce D3-branes wrapping supersymmetry-preserving cycles, and we focus on cycles with an odd number of directions on either factor of the ten-dimensional solution. We show that the D3-brane action can be naturally written in terms of the gauge field and bilinears constructed out of the supersymmetry spinors, and can then be evaluated using an odd-dimensional application of the Berline--Vergne--Atiyah--Bott equivariant localization theorems \cite{Goertsches:2015vga}.

As a concrete example, we look at the supersymmetric Kerr--Newman-AdS$_5$ black hole with unequal angular momenta and equal electric charges introduced in \cite{Chong:2005hr}. Its on-shell action, computed via a complex supersymmetry-preserving deformation, reproduces the $\cO(N^2)$ term in the large $N$ limit of the superconformal index of the dual four-dimensional $\cN=1$ superconformal field theory \cite{Cabo-Bizet:2018ehj}, where the choice of field theory is encoded in the choice of internal Sasaki--Einstein five-manifold. We introduce probe D3-branes wrapping various cycles in the ten-dimensional geometry, and compute their action using equivariant localization.
We first consider D3-branes wrapping a circle on the horizon of the black hole and a three-cycle in the internal space, which represent non-perturbative corrections of order $\ex^{-N}$ to the Bethe Ansatz expansion of the superconformal index \cite{Aharony:2021zkr}. We next consider D3-branes wrapping the horizon of the black hole and a circle in the internal space, whose physical interpretation remains less clear \cite{Aharony:2021zkr}. Finally, we study branes wrapping a non-compact three-cycle inside the black hole and a circle in the internal space. These are holographically dual to the insertion of a BPS superconformal defect in the dual field theory, and their action represents the correction at order $\cO(N)$ to the four-dimensional superconformal index arising from the inclusion of the two-dimensional superconformal index of the defect fields \cite{Chen:2023lzq}.\footnote{We emphasize that these corrections are distinct from the corrections obtained by considering higher derivative terms in the action \cite{Bobev:2022bjm, Cassani:2022lrk}, which have recently been computed using localization \cite{Gaar:2026nqq}.}

The corresponding brane actions were evaluated by explicit computation in the $S^5$ case, corresponding to the $4d$ $\cN=4$ SYM field theory, in \cite{Aharony:2021zkr, Chen:2023lzq}. In this note, we show that the simplicity of the expressions obtained there can be straightforwardly explained and reproduced using equivariant localization, and we extend them to the case of arbitrary toric Sasaki--Einstein internal space. 


\section{IIB geometry}

We begin by describing the ten-dimensional supersymmetric background obtained by uplifting the Kerr--Newman-AdS$_5$ black hole on a Sasaki--Einstein five-manifold.


\subsection{Five-dimensional minimal gauged supergravity}

Later in the note, we will focus on a complex supersymmetry-preserving deformation of the supersymmetric Kerr--Newman-AdS$_5$ black hole \cite{Chong:2005hr}, which captures the holographic dual to the superconformal index of four-dimensional $\cN=1$ superconformal field theories \cite{Cabo-Bizet:2018ehj, GonzalezLezcano:2019nca, Lanir:2019abx, Benini:2020gjh}. It is an example of a supersymmetric solution of five-dimensional minimal gauged supergravity, which (at the bosonic level) describes the interactions of the metric and the U$(1)$ gauge field $A$ with field strength $F \equiv \dd A$ according to the action
\begin{equation}\label{eq:EMaction}
S_5 = \frac{1}{16\pi G_{(5)}} \int \Big[ \big( R + 12\big)\vol - \tfrac{3}{2} \,  F\wedge \star_5 F  -A\wedge  F \wedge F \Big] \, .
\end{equation}
The existence of a supersymmetric solution can be rephrased in terms of torsion conditions on differential forms constructed from the Killing spinor \cite{Gauntlett:2003fk}.
In particular, the rank-one bilinear is dual to a non-spacelike Killing vector field $V$ which also preserves the gauge field. Focusing on solutions for which $V$ is timelike, with $V_aV^a \equiv - f^2$, we can write the five-dimensional metric on $M_5$ as a circle foliation over a base space $M_4$
\begin{equation}
\label{eq:Metric_AAdS5}
\dd s^2(M_5) = -f^2\sigma_0^2+ f^{-1}\dd s^2(M_4) \, , 
\end{equation}
where $\sigma_0 \equiv - f^{-2} V^\flat$ is by construction a globally well-defined one-form, and $\dd s^2(M_4)$ is a K\"ahler metric on the base space, with K\"ahler form denoted by $X$.

Provided $M_5$ admits at least one further commuting isometry, the on-shell action from \eqref{eq:EMaction} can be computed using equivariant localization \cite{Colombo:2025ihp, Colombo:2025yqy, BenettiGenolini:2025icr}. 
A key ingredient in performing the localization is the connection
\begin{equation}
\label{eq:5d_Form_sigma}
    \sigma=f \sigma_0+A\, .
\end{equation} 
Note that, because of the presence of the gauge field, $\sigma$ may not be a global one-form, and for a general supersymmetric solution one has to define its integral in patches.
The field strength $F$ is uniquely fixed by supersymmetry \cite{Gauntlett:2003fk}, and is the top form of an equivariantly closed form, since $V \hook F = \dd f$. Choosing a gauge such that $\cL_V A = 0$, we find that $V \hook A = - f + \alpha$, where $\alpha$ is an arbitrary constant defined patchwise, and $V \hook \sigma = \alpha$.
A natural choice is to fix $\alpha=0$ \cite{Gauntlett:2003fk, BenettiGenolini:2025icr}, but then the gauge field would be singular where $V=0$. Instead, here we keep $\alpha$ non-zero. 
Note that when we restrict to the supersymmetric black hole, $A$ can be defined globally, and similarly $\sigma$.
To make sure that $A$ is regular on the horizon, we fix a non-zero $\alpha$ that can be identified as being proportional to a chemical potential conjugate to the electric charge \cite{Cabo-Bizet:2018ehj, Colombo:2025yqy}.


\subsection{Uplift on a Sasaki--Einstein five-manifold }
\label{subsec:Uplift_SE5}

Any supersymmetric solution of five-dimensional minimal gauged supergravity can be locally uplifted to a solution of type IIB supergravity on a five-dimensional Sasaki--Einstein manifold, denoted by SE$_5$ \cite{Buchel:2006gb}.
To describe the uplift, it is convenient to express the Sasaki--Einstein manifold in the canonical form of a U$(1)$ foliation over a K\"ahler--Einstein base $\mathcal{B}_4$. On the base we take the K\"ahler form to be $J$, and the Ricci-form potential to be $P$, which satisfies $\dd P=6 J$. The Killing vector generating the foliation, known as the Reeb vector field, is denoted by $\xi$. Its dual one-form is denoted by $\eta$, and satisfies
\begin{equation}
\label{eq:SE5_ContactStructure}
   \xi\hook \eta=1\, ,\quad \xi\hook \dd\eta=0\, ,\quad \dd \eta=2J\, .
\end{equation}
In particular, $\dd \eta$ is basic with respect to $\xi$, and $\eta$ defines a contact structure. A Sasaki--Einstein manifold admits a Killing spinor $\chi$ from which one may construct $\eta$ and $J$ as spinor bilinears. 
Finally, we can write the metric on SE$_5$ as a U$(1)$ foliation over the base
\begin{equation}
    \dd s^2(\mathrm{SE}_5) = \eta^2 + \dd s^2(\mathcal{B}_4)\, .
\end{equation}

To perform the uplift, we introduce $\eta_A=\eta-A$.
The ten-dimensional metric is supported by the self-dual five-form Ramond--Ramond flux only
 \begin{equation}
 \label{eq:10d_Split_Metric}
 \begin{split}
 \dd s^2_{(10)}&=\dd s^2(M_5)+\eta_A^2+\dd s^2(\mathcal{B}_4)\, ,\\
\cF&=\dd \cC\equiv\dd \mathcal{C}^{\text{bulk}}+\dd \mathcal{C}^{A}\, .
 \end{split}
 \end{equation}
Here, we highlighted that the four-form potential splits in terms of a universal bulk piece, independent of the gauge field, and a second piece involving the gauge field
 \begin{equation}\label{eq:Csplit}
 \begin{split}
\cC^{\text{bulk}}&=-4 \, \Xi_4 -\frac{1}{3} \eta\wedge P\wedge J\, ,\\
\cC^A&=-\frac{1}{2}\eta \wedge(2 A \wedge J+\star_{5}F+A\wedge F)\, ,
 \end{split}
 \end{equation}
 where $\dd \Xi_4=\vol(M_5)$ and similarly the exterior derivative of the second term in $\cC^{\text{bulk}}$ gives $\vol({\rm SE}_5)$.


\section{Probe D3-branes}

We are interested in computing the action of supersymmetric probe D3-branes in the ten-dimensional background just described. 
The effective theory of a D3-brane wrapped on a four-dimensional cycle $\Sigma$ is given by the following DBI and Chern--Simons action \footnote{Note that we have set the gauge-invariant worldvolume field strength on the brane to vanish since it plays no role in the following.}
\begin{equation}\label{eq:action}
        S_{\mathrm{D3}}=-\mu_\mathrm{D3}\int_\Sigma \Big[\sqrt{-\det(g|_\Sigma)}\dd^{4}x- \mathcal{C}\Big]\,, \quad  
    \mu_\mathrm{D3}^{-1}=(2\pi)^3\ls^4\,.
\end{equation}
In order for the probe brane to preserve supersymmetry, the $\kappa$-symmetry condition must be satisfied, imposing that $\Sigma$ is a generalized calibrated cycle.
Using generalized geometry, the $\kappa$-symmetry condition for a probe D3-brane can be reformulated in terms of a set of ten-dimensional spinor bilinears, constructed using the two Majorana--Weyl spinors $\varepsilon_1,\varepsilon_2$ of the ten-dimensional solution \cite{Martucci:2011dn}. In particular, on the worldvolume of a calibrated D3-brane one has that 
\begin{equation}\label{eq:calib}
     [(\dd z^M \wedge \Phi)|_\Sigma]\big|_{4}= K^{M}\sqrt{-\det(g|_\Sigma)}~\dd^{4}x\, ,
\end{equation}
where $z^M$ are the ten-dimensional coordinates, $K$ is the ten-dimensional Killing vector 
\begin{equation}\label{eq:Kdef}
    K=\frac{1}{2}(\bar{\varepsilon}_1\Gamma^{M}\varepsilon_1+\bar{\varepsilon}_2\Gamma^{M}\varepsilon_2) \, \partial_{M}\,,
\end{equation}
and $\Phi$ is the polyform obtained via the Clifford map from the bispinor 
\begin{equation}\label{eq:Phidef}
    \Phi = 32 \, \varepsilon_2 \otimes \bar{\varepsilon}_1 \, .
\end{equation}
The key point of this reformulation is that the calibration condition allows  us to rewrite the DBI part of the action using spinor bilinears. 
We observe that the Killing spinor $\varepsilon \equiv \varepsilon_1 + \ii \varepsilon_2$ is uncharged under $K$: this will allow us to fix the weights of the Killing vector later. 

In analogy with the fibration form of the metric \eqref{eq:10d_Split_Metric}, the ten-dimensional Clifford algebra can be split in terms of the Clifford algebras on the two five-dimensional factors, and the same can be done for the ten-dimensional Killing spinor $\varepsilon$ (see the supplementary material for more details).
Accordingly, the ten-dimensional Killing vector has the expression
\begin{equation}\label{eq:Kin5}
    K = \frac{1}{2}(V+ \alpha \xi)\, ,
\end{equation}
which explicitly depends on the parameter $\alpha$, though note that $K$ is null for all values of $\alpha$. 
We can similarly proceed to compute the expression of the three-form part of the polyform $\Phi$ in terms of the spinor bilinears in the five-dimensional factors.
We focus on probe D3-branes wrapping cycles with either a $1+3$ or a $3+1$ split of legs across the two factors of the ten-dimensional space. Accordingly, in the expansion of $\Phi_3$ we keep only the terms of this type, and find that the relevant contributions are
\begin{equation}\label{eq:Phi3in5}
    \Phi_3 = \frac{1}{2}(\sigma-\eta)\wedge \big(f J-X)\, .
\end{equation}
Remarkably, the one-form $\sigma$ introduced in \eqref{eq:5d_Form_sigma} as key for performing localization in 5$d$ naturally emerges from the ten-dimensional analysis!

We can now evaluate the action $S_{\text{D}3}$ in \eqref{eq:action} by inserting the expression for $\mathcal{C}$ in \eqref{eq:Csplit} and the calibration condition \eqref{eq:calib} with $\Phi$ as in \eqref{eq:Phi3in5}. Then, for the cases we are interested in, the action is given by the remarkably simple formula
\begin{equation}\label{eq:Sd3final}
    S_\mathrm{D3}=\frac{\mu_\mathrm{D3}}{2}\int_{\Sigma}\Big[\sigma\wedge \eta\wedge \dd\eta +\sigma\wedge \dd \sigma\wedge \eta\Big]\, .
\end{equation}
Note that by construction, only one of the two terms will be non-vanishing.

We have now derived \eqref{eq:Sd3final}, the action of a probe D3-brane on the ten-dimensional geometry constructed by uplifting a general supersymmetric solution of $5d$ minimal gauged supergravity $M_5$ on an arbitrary SE$_5$. The final expression has a particularly simple form, neatly written in terms of spinor bilinears on $M_5$ and SE$_5$, and the gauge field $A$.
In the remainder of this note, we will use equivariant localization to evaluate \eqref{eq:Sd3final} in the case where $M_5$ is the Kerr--Newman-AdS$_5$ black hole, while SE$_5$ remains an arbitrary toric Sasaki--Einstein manifold. This setup allows us to probe the superconformal index of a large family of four-dimensional quiver superconformal field theories. That said, the formula \eqref{eq:Sd3final} has a significantly larger range of applicability, which will be explored in \cite{toappear}.


\section{Elements of localization}
 
In order to evaluate \eqref{eq:Sd3final} on the relevant cycles, it is useful to review some aspects of equivariant localization for odd-dimensional toric manifolds as developed in \cite{Goertsches:2015vga}. For the purposes of this note only a small number of formulas will be needed, and we refer to \cite{Goertsches:2015vga, Colombo:2025ihp,Colombo:2025yqy, toappear} for more details. 
We proceed in stages of increasing specialization. We first present general expressions for equivariant integrals in five dimensions in terms of toric data, then review their physical interpretation when applied to the internal SE$_5$, and finally we summarize the toric data relevant for the Kerr--Newman-AdS$_5$ black hole.

Before doing so, let us recall that the localization results used here apply strictly to Riemannian manifolds, whereas our discussion so far has been formulated in Lorentzian signature, cf.\ \eqref{eq:EMaction}.
It would be interesting to carry out the analysis entirely in Euclidean signature, as in \cite{BenettiGenolini:2025icr}, but here we will work with an analytic continuation of the Lorentzian supersymmetric solution \eqref{eq:Metric_AAdS5}, as in \cite{Cabo-Bizet:2018ehj, Colombo:2025ihp,Colombo:2025yqy}.

\subsection{Localization on toric \texorpdfstring{$5$}{5}-manifolds}

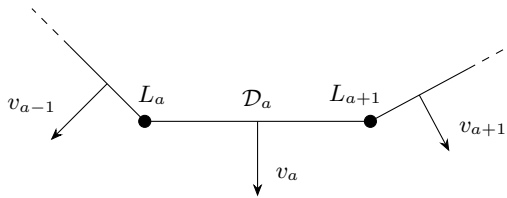
\begin{figure}
    \begin{center}
    \begin{tikzpicture}

    \draw[] (-2.5,1.)--(-1.5,0)--(1.5,0)--(2.8,.7);
    \draw[dashed] (-2.5,1.)--(-3,1.5);
    \draw[dashed] (2.8,.7)--(3.35,1.);
    
    \filldraw[] (1.5,0) circle (.08);
    \node at (1.3,0.35) {$L_{a+1}$};
    \node at (0.,0.3) {$\cD_a$};
    \draw[-Stealth] (0,0)--(0,-1.);
    \node at (0.4,-.7) {$v_a$};

    \filldraw[] (-1.5,0) circle (.08);
    \node at (-1.4,0.35) {$L_a$};
    
    \draw[-Stealth] (-2,.5)--(-2.75,-.25);
    \node at (-3,.2) {$v_{a-1}$};
   
    \draw[-Stealth] (2.15,.35)--(2.55,-.4);
    \node at (3,-.1) {$v_{a+1}$};
        
    \end{tikzpicture}

    \end{center}

\caption{A subset of a generic toric diagram. The edges, defined by the vectors $v_a$, are the loci where a circle shrinks and define three-cycles $\cD_a$. The vertices, denoted by $L_a$ are known as leaves.
    \label{fig:toric}}
\end{figure}

We consider a five-dimensional toric manifold $\cM$, not necessarily compact. We view the manifold as a U$(1)^3$ fibration over a two-dimensional polytope (toric diagram) which closes if the space is compact but is open if it is non-compact. The faces of the polytope are the loci where a circle inside U$(1)^3$ degenerates and are labelled by a three-component vector $v_a$, see figure \ref{fig:toric}. Such faces represent three-dimensional sub-manifolds, $\cD_a$ where the isotropy group is U$(1)$. Neighbouring faces are ordered sequentially. The intersection of two such faces defines a leaf $\cD_{a-1}\cap \cD_{a}=L_a$, which is isomorphic to a circle, whilst the intersection of non-sequential faces is empty.

We equip the toric manifold with a Killing vector $\kv$ and introduce a globally well-defined one-form $\cf$ such that $\dd \cf$ is basic with respect to $\kv$, i.e.\ $\kv \hook \dd \cf=0$. 
For the five-dimensional spaces we consider, both the Killing vector and the one-form are guaranteed by supersymmetry.\footnote{In general, the one-form $\cf$ needs only be well-defined in a patch covering the leaf, and one must carefully track the U$(1)^3$-transformations between patches, as in \cite{Colombo:2025yqy}. For our applications to the black hole, $\cf$ is globally well-defined and so we can ignore this subtlety and work with a globally well-defined one-form.} 
Assuming further that $\cf$ is invariant under $\kv$, $\kv\hook \cf$ is constant on the leaves.
In the following we are interested in computing integrands of the form $\cf$ or $\cf\wedge \dd \cf$ and so we will restrict to these, though the formula may be extended to more general integrands.
The integral of $\cf$ over a leaf $L_a$ can be written in terms of the toric data as
\begin{equation}
\label{eq:Leaf_Integrals}
    \int_{L_a}\cf = 2\pi\frac{(\kv\hook \cf|_{L_a})}{(\kv,v_{a-1},v_a)}\, ,
\end{equation}
where we have used the shorthand notation $\det(\kv,v_{a-1},v_a)\equiv(\kv,v_{a-1},v_a)$. 
On the other hand, the integral of $\cf\wedge \dd\cf$ over a compact three-cycle $\cD_a$, i.e.\ represented by a finite face of the toric polytope, can be written as \footnote{This expression is considerably simplified by the assumption that $\cf$ is globally well-defined. When it is necessary to introduce patches, one also receives contributions from the interfaces between the patches, which are not relevant for our applications.}
\begin{align}\label{eq:3dloc}
\int_{\cD_a}\cf\wedge\dd \cf &=(2\pi)^2 \frac{(\kv\hook \cf |_{L_a})^2 (v_{a-1},v_a,v_{a+1})}{(\kv,v_{a-1},v_{a})(\kv,v_{a},v_{a+1})}\, .
\end{align}
For a non-compact three-cycle this needs a little refinement that we will address later.


\subsection{A short \texorpdfstring{SE$_5$}{SE5} digression}

We now consider the case of a toric Sasaki--Einstein five-manifold, and label the faces of the toric diagram by $d$ outward-pointing normal vectors $v_I$, $I=1,...,d$. As described in section \ref{subsec:Uplift_SE5}, this space satisfies the assumptions in the previous discussion, because the spinor bilinears $\xi$ and $\eta$ are globally defined and satisfy \eqref{eq:SE5_ContactStructure}.
We introduce generators of the U$(1)^3$ action $\partial_{\phi_i}$, where $i=1,2,3$, and each $\phi_i$ is $2\pi$-periodic. We further impose that the Killing spinor has unit charge under $\partial_{\phi_1}$ but is not charged under either of $\partial_{\phi_2}$ or $\partial_{\phi_3}$. This is a convenient choice of basis, because it guarantees that the first entry of each $v_I$ is $1$.
We may expand the Reeb vector field in this basis as
\begin{equation}
\label{eq:SE5_xi}
    \xi=\sum_{i=1}^{3} b_{i} \, \partial_{\phi_i}\, .
\end{equation}
The relevance of toric SE$_5$ to us is that type IIB supergravity backgrounds of the form AdS$_5\times {\rm SE}_5$ are holographically dual to four-dimensional $\cN=1$ superconformal field theories living on a stack of $N$ D3-branes sitting at the apex of the cone over SE$_5$. The $a$-central charge and the flux integer $N$ are related to the the geometry by
\begin{equation}
    a_\mathrm{CFT}=\frac{ \pi^3}{4 \, \vol(\text{SE}_5)}N^2\, ,\qquad N=\frac{4 \, \vol(\mathrm{SE}_5)}{(2\pi\ls)^4}\, ,
\end{equation}
where
\begin{equation}
    \vol(\text{SE}_5)=\frac{1}{4}\int_{\text{SE}_5} \eta\wedge \dd \eta \wedge \dd \eta\, .
\end{equation}
This can be computed via localization using the Sasakian volume formula \cite{Martelli:2005tp,Martelli:2006yb}. Moreover, supersymmetry of the solution fixes the coefficient $b_1$ in \eqref{eq:SE5_xi} to be $b_1=3$, whereas the remaining two parameters can be fixed by $a$-maximisation \cite{Martelli:2005tp,Martelli:2006yb}.

As described in the previous section, each face of the toric polytope corresponds to a three-cycle, which for the SE$_5$ we denote by $\Sigma_I$.
The contact volume of these three-cycles, given by \cite{Berenstein:2002ke}
\begin{equation}
    \vol(\Sigma_I)=\frac{1}{2}\int_{\Sigma_I}\eta \wedge \dd\eta\, ,
\end{equation}
is proportional to the R-charge of BPS operators in the dual field theory associated to a probe D3-brane wrapped on the corresponding three-cycle
\begin{equation}
    \Delta_I = \frac{\pi}{3} \frac{\vol(\Sigma_I)}{\vol(\mathrm{SE}_5)} \, , \qquad \sum_{I=1}^{d} \Delta_I=2\, .
\end{equation}
These can be computed using \eqref{eq:3dloc}, with the further simplification that $\xi \hook \eta =1$ everywhere on SE$_5$.


\begin{figure}
    \centering
    \begin{tikzpicture}

\begin{scope}

    \draw[] (-1.5,0)--(0,0)--(-1.5,1.5)--(-1.5,0);

    \filldraw[] (0,0) circle (.08);
    \filldraw[] (-1.5,0) circle (.08);
    \filldraw[] (-1.5,1.5) circle (.08);

    \node at (-1.7,-0.2) {$L_3$};
    \node at (.3,-0.2) {$L_1$};
    \node at (-1.1,1.5) {$L_2$};

    \node at (-0.5,2) {$S^5$};

    \draw[-Stealth] (-.75,0)--(-.75,-.8);
    \node at (-0.5,-.6) {$v_3$};

    \draw[-Stealth] (-1.5,.75)--(-2.3,.75);
    \node at (-2,1.) {$v_2$};

    \draw[-Stealth] (-.75,.75)--(-0.25,1.25);
    \node at (-.3,.85) {$v_1$};

    \end{scope}

    \begin{scope}[xshift=3 cm,yshift=.5cm]
        
    \draw[] (-.6,-.6)--(.6,-.6)--(.6,.6)--(-.6,.6)--(-.6,-.6);

    \filldraw[] (-.6,-.6) circle (.08);
    \filldraw[] (-.6,.6) circle (.08);
    \filldraw[] (.6,-.6) circle (.08);
    \filldraw[] (.6,.6) circle (.08);

    \node at (-.8,-.8) {$L_1$};
    \node at (-.85,.85) {$L_4$};
    \node at (.85,-.8) {$L_2$};
    \node at (.8,.85) {$L_3$};

    \node at (0,1.5) {$T^{1,1}$};

    \draw[-Stealth] (-.6,0)--(-1.2,.0);
    \node at (-1,.2) {$v_4$};
    \draw[-Stealth] (0,-.6)--(0,-1.2);
    \node at (.3,-1) {$v_1$};
    \draw[-Stealth] (.6,0)--(1.2,.0);
    \node at (1,.2) {$v_2$};
    \draw[-Stealth] (0,.6)--(0,1.2);
    \node at (.3,1) {$v_3$};

    \end{scope}

    \end{tikzpicture}

    \caption{Toric diagrams of $S^5$ and $T^{1,1}$. The diagrams are a projection of a $3d$ polytope to the plane. }
    \label{fig:S5andT11}
\end{figure}
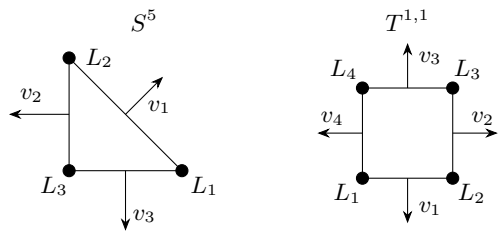

For concreteness, we will evaluate the general integrals for two choices of SE$_5$. First, the round $S^5$, with toric vectors
\begin{equation}
    v_1=(1,0,0)\,,\quad v_2=(1,1,0)\,, \quad v_3=(1,0,1) \, .
\end{equation}
The toric diagram is represented in figure \ref{fig:S5andT11}, and the on-shell Reeb vector is $\xi=(3,1,1)$, so that the leaf integrals \eqref{eq:Leaf_Integrals} are all $2\pi$.
The dual field theory is $\cN=4$ SYM; the $a$-central charge and R-charges of BPS operators are given by
\begin{equation}
    a_{\text{SYM}}=\frac{N^2}{4}\,, \quad \Delta_I=\frac{2}{3}\,,
\end{equation}
for each $I=1,2,3$.  
The extension to a quotient of $S^5$ by a discrete subgroup of $SO(4)$ is straightforward.

Second, we consider SE$_5 = T^{1,1}$ \cite{Candelas:1989js}. The toric vectors are 
\begin{equation}
\begin{aligned}
v_1 &=(1,1,1)\, ,&\quad v_2&=(1,0,1)\, , \\
v_3 &=(1,0,0)\, , &\quad v_4&=(1,1,0)\, ,
\end{aligned}
\end{equation} see figure \ref{fig:S5andT11}, and the on-shell Reeb vector is $\xi = (3,3/2,3/2)$. The dual field theory, living on D3-branes at a conifold singularity, is the Klebanov--Witten theory \cite{Klebanov:1998hh}. The $a$-central charge and R-charges of BPS operators for each $I=1,2,3,4$ are given by
\begin{equation}
    a_{\text{KW}}=\frac{27}{64} N^2 \,, \quad \Delta_I=\frac{1}{2}\,.
\end{equation}


\subsection{The \texorpdfstring{Kerr--Newman-AdS$_5$}{Kerr-Newman-AdS5} black hole}

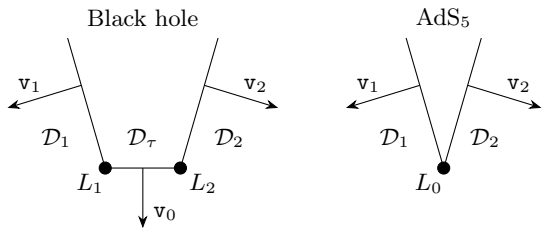
\begin{figure}
    \centering
    \begin{tikzpicture}

\begin{scope}[xshift=4.5 cm]

    \draw[] (-1.5,1.73)--(-1,0)--(0,0)--(.5,1.73);

    \filldraw[] (0,0) circle (.08);
    \filldraw[] (-1.,0) circle (.08);

    \node at (-1.2,-0.2) {$L_1$};
    \node at (.3,-0.2) {$L_2$};

    \node at (-1.65,0.4) {$\cD_1$};
    \node at (.65,0.4) {$\cD_2$};
    \node at (-.5,0.4) {$\cD_\tau$};
    \node at (-0.5,2) {Black hole};

    \draw[-Stealth] (-.5,0)--(-.5,-.8);
    \node at (-0.2,-.6) {$\vBH_0$};

    \draw[-Stealth] (-1.32,1.1)--(-2.3,.8);
    \node at (-2,1.1) {$\vBH_1$};

    \draw[-Stealth] (.32,1.1)--(1.3,.8);
    \node at (1,1.1) {$\vBH_2$};

    \end{scope}

    \begin{scope}[xshift=9 cm]
        
    \draw[] (-1.5,1.73)--(-1,0)--(-0.5,1.73);

    \filldraw[] (-1.,0) circle (.08);

    \node at (-1.2,-0.2) {$L_0$};

    \node at (-1.65,0.4) {$\cD_1$};
    \node at (-.45,0.4) {$\cD_2$};
    \node at (-1,2) {$\text{AdS}_5$};

    \draw[-Stealth] (-1.32,1.1)--(-2.3,.8);
    \node at (-2,1.1) {$\vBH_1$};

    \draw[-Stealth] (.32-1,1.1)--(.3,.8);
    \node at (0,1.1) {$\vBH_2$};

    \end{scope}

    \end{tikzpicture}

    \caption{Toric diagrams of the Kerr--Newman-AdS$_5$ black hole and thermal AdS$_5$.}
    \label{fig:BHandAdStoric}
\end{figure}

Finally, we describe the toric geometry of the asymptotically AdS$_5$ space $M_5$. 

We are interested in studying the Kerr--Newman-AdS$_5$ black holes obtained by analytically continuing the supersymmetric Lorentzian solutions in \cite{Chong:2005hr}
which may be described by the toric vectors
\begin{equation}
  \vBH_0=(1,0,0)\, ,\quad   \vBH_1=(0,0,1)\, ,\quad \vBH_2=(0,1,0)\, ,
\end{equation}
see figure \ref{fig:BHandAdStoric}. 
The conformal boundary is $S^1\times S^3$ and the asymptotic circle shrinks along the face defined by $\vBH_0$. The black hole horizon in the Wick-rotated geometry becomes the three-cycle $\cD_{\tau}$ depicted in figure \ref{fig:BHandAdStoric}, which is diffeomorphic to $S^3$. The two three-cycles $\cD_1$ and $\cD_2$ are both homeomorphic to $\mathbb{R}^2\times S^1$, albeit with different circles. The supersymmetric Killing vector $V$ is
\begin{equation}\label{eq:BHV}
    V=\frac{1}{\beta}\big(2\pi \ii \, \vBH_0-\omega_1 \vBH_2-\omega_2 \vBH_1)\, ,
\end{equation}
where $\beta$ is the inverse temperature. We are working in conventions in which the degenerating vector fields defined by the $\vBH$'s are $2\pi$-periodic. The parameters $\omega_1$ and $\omega_2$ are the complex chemical potentials conjugate to the angular momenta of the black hole, and we have exchanged the ordering in $V$ for later convenience.  Regularity of the gauge field at the horizon fixes the constant $\alpha$, which via UV-IR relations turns out to be equal to the boundary potential conjugate to the electric charge $\varphi = \alpha \beta$ \cite{Cabo-Bizet:2018ehj, Colombo:2025yqy} 
\begin{equation}
    V\hook\sigma= \frac{\varphi}{\beta}\,.
\end{equation} 
Moreover, one finds that the weights and $\varphi$ are constrained by 
\begin{equation}
\label{eq:Constraint_ChemicalPotentials}
    2\pi\ii = \omega_1 + \omega_2 - 3 \varphi \, .
\end{equation}
This can be derived either using the methods of \cite{Cabo-Bizet:2018ehj, Colombo:2025ihp} or by imposing that the ten-dimensional Killing spinor is uncharged under $K$. 

The on-shell action for a general toric Sasaki--Einstein uplift of the black hole can also be computed using equivariant localization and one finds \cite{Cabo-Bizet:2018ehj}
\begin{equation}
    I= 2 \frac{\varphi^3}{\omega_1\omega_2} a_{\text{CFT}}\,.
\end{equation}
We emphasize that deriving this result does not require the analytic expression for the metric of the black hole, but only the toric data and the UV-IR relation.\footnote{For reference, the map between the parameters here 
and in \cite{Aharony:2021zkr, Chen:2023lzq} is $(\omega_{1}, \omega_2, \varphi)^{\rm here} = 2\pi\ii (\sigma, \tau, \Delta)^{\rm there}$. In the language of \cite{Aharony:2021zkr}, we are on the ``first branch'' of solutions. There is also a ``second branch'' of solutions where the left-hand side of \eqref{eq:Constraint_ChemicalPotentials} is $-2\pi \ii$.}


\section{Brane actions}\label{sec:onshell}

We can now use our earlier results to compute the action of probe D3-branes wrapping cycles in the ten-dimensional geometry obtained by uplifting the Kerr--Newman-AdS$_5$ black hole on an arbitrary SE$_5$. In particular, we consider three families of probe branes wrapping different cycles. For each case, after discussing the case of arbitrary SE$_5$, we specialize to $S^5$ and $T^{1,1}$, and for concreteness we represent the brane on the toric diagram corresponding to black hole$\times S^5$.

\subsection{Non-perturbative corrections to the 
index}

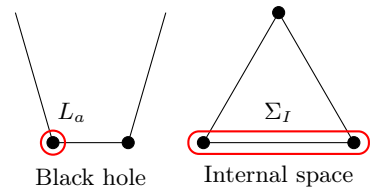
\begin{figure}
    \centering
    \begin{tikzpicture}

  \begin{scope}[xshift=8cm]
        
    \draw[] (-1.5,1.73)--(-1,0)--(0,0)--(.5,1.73);

    \filldraw[] (0,0) circle (.08);
    \filldraw[] (-1.,0) circle (.08);
    \draw[thick,red] (-1.,0) circle (0.16);

    \node at (-.75,0.4) {$L_a$};
    \node at (-0.5,-0.45) {Black hole};

    \end{scope}

    \begin{scope}[xshift=10 cm]
        
    \draw[] (-1,0)--(1,0)--(0,1.73)--(-1,0);

    \filldraw[] (1.,0) circle (.08);
    \filldraw[] (-1.,0) circle (.08);
    \filldraw[] (0,1.73) circle (.08);

    \draw[rounded corners=4pt, draw=red, thick] (-1.2,-.15) rectangle (1.2,.15);

    \node at (0,0.4) {$\Sigma_I$};
    \node at (0,-0.45) {Internal space};

    \end{scope}

\end{tikzpicture}

    \caption{The brane wraps a three-cycle in the internal space and one of the two leaves of the black hole. }
    \label{fig:1+3wrapping}
\end{figure}

One method for evaluating the supersymmetric index is known as the Bethe Ansatz method. This expresses the index as a sum over solutions to the  Bethe Ansatz equations. Every solution to the Bethe Ansatz equations that admits a good large $N$ limit may be identified with a specific Euclidean solution on the gravity side: either a Euclidean black hole solution or an orbifold thereof. In addition to the contribution of order $\cO(N^2)$, there are non-perturbative corrections which, on the gravity side, arise from D3-branes wrapping three-cycles inside the Sasaki--Einstein and a circle in the black hole \cite{Aharony:2021zkr}.

Three-cycles which the brane may wrap inside the Sasaki--Einstein manifold are given by the $d$ faces of the toric diagram of the internal space. On the other hand, the circle that the brane may wrap inside the black hole is given by one of the leaves, i.e.\ the intersection of the two faces. In total there are $2 \times d$ inequivalent ways of wrapping the probe D3-branes in this case, and fortunately they can all be computed analogously. For $S^5$, we have represented this pictorially in figure \ref{fig:1+3wrapping}.

Looking at the equation \eqref{eq:Sd3final} for the action, we see that only the first term has support on the brane worldvolume, and we find
\begin{equation}
\begin{split}\label{eq:S1S5gen}
    S_{\text{D}3}(L_a\times \Sigma_I)&= \frac{\mu_{\text{D3}}}{2}\int_{L_a}\sigma\int_{\Sigma_I}\eta\wedge \dd \eta\\
    &=3\pi N\Delta_I \frac{\varphi}{\omega_a}\, ,
    \end{split}
\end{equation}
for $a=1,2$ and $I=1,\dots,d$. 
Choosing the SE$_5$ to be $S^5$ gives
\begin{equation}
    S_{\text{D}3}^{S^5}(L_a\times \Sigma_I) = 2\pi N \frac{ \varphi}{\omega_a}\, ,
\end{equation}
which agrees with the result in \cite{Aharony:2021zkr}. On the other hand, for $T^{1,1}$
\begin{equation}
    S_{\text{D}3}^{T^{1,1}}(L_a\times \Sigma_I) = \frac{3\pi}{2} N \frac{ \varphi}{\omega_a}\, .
\end{equation}
Equation \eqref{eq:S1S5gen} is a prediction for the non-perturbative corrections to the index, arising from the Bethe Ansatz method, for a large class of quiver $4d$ superconformal field theories. It would be interesting to compute these directly from the field theory, for instance by applying the techniques used in \cite{Aharony:2021zkr} for $\mathcal{N}=4$ SYM to the results in \cite{Benini:2020gjh}.


\subsection{Wrapping the horizon}

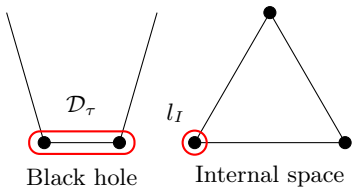
\begin{figure}
    \centering
    \begin{tikzpicture}

    \begin{scope}[xshift=0 cm]
        
    \draw[] (-1.5,1.73)--(-1,0)--(0,0)--(.5,1.73);

    \filldraw[] (0.,0) circle (.08);
    \filldraw[] (-1.,0) circle (.08);

    \draw[rounded corners=4pt, draw=red, thick] (-1.2,-.15) rectangle (.2,.15);

    \node at (-0.5,0.5) {$\cD_\tau$};
    \node at (-0.5,-0.45) {Black hole};

    \end{scope}

    \begin{scope}[xshift=2 cm]
        
    \draw[] (-1,0)--(1,0)--(0,1.73)--(-1,0);

    \filldraw[] (1.,0) circle (.08);
    \filldraw[] (-1.,0) circle (.08);
    \filldraw[] (0,1.73) circle (.08);
    \draw[thick,red] (-1.,0) circle (0.16);

    \node at (-1.25,0.4) {$l_I$};
    \node at (0,-0.45) {Internal space};

    \end{scope}
    \end{tikzpicture}

    \caption{The probe brane is wrapped on the horizon and a circle inside the internal space.}
    \label{fig:3+1wrapping}
\end{figure}
A second option is to wrap the $S^3$ inside the black hole and one of the circles of the SE$_5$, which is necessarily a leaf.
The interpretation of these probe D3-branes is more mysterious than those considered in the previous section; however, they would still contribute to the gravitational path integral \cite{Aharony:2021zkr}. Since there is a unique such three-sphere inside the black hole and $d$ independent leaves for the SE$_5$, there are $d$ inequivalent configurations. We have represented this pictorially for an $S^5$ internal space in figure \ref{fig:3+1wrapping}.

The probe brane action is
\begin{equation}
\label{eq:D3_probe_Mysterious}
    \begin{split}
        S_{\text{D3}}(\mathcal{D}_{\tau}\times l_I)&=\frac{\mu_\mathrm{D3}}{2}\int_{\mathcal{D}_{\tau}}\sigma\wedge \dd \sigma \int_{l_I}\eta\\
        &=\frac{a_{\text{CFT}}}{N} \frac{8 \pi}{(\xi, v_{I-1},v_{I})}\frac{\varphi^2}{\omega_1\omega_2}\, ,
    \end{split}
\end{equation}
which is a new result for a generic toric Sasaki--Einstein manifold. 
Specializing to the $S^5$ we find
\begin{equation}
\label{eq:D3Antibrane_S5}
    S_{\text{D3}}^{S^5}(\mathcal{D}_{\tau}\times l_I) = 2\pi N \frac{\varphi^2}{\omega_1\omega_2} \,, 
\end{equation}
in agreement with \cite{Aharony:2021zkr}.\footnote{As written, the result \eqref{eq:D3Antibrane_S5} is the action of a D3-brane, which is in fact only supersymmetric on the ``second branch,'' where the parameters satisfy $\omega_1 + \omega_2 - 3\varphi = - 2\pi \ii$ (in contrast to \eqref{eq:Constraint_ChemicalPotentials}) \cite{Aharony:2021zkr}. On the first branch, only anti-D3-branes are supersymmetric. By construction, this changes the sign of the WZ term in \eqref{eq:action}, but also the sign in the $\kappa$ symmetry condition, and hence the sign in the calibration condition \eqref{eq:calib} (see \cite{Martucci:2011dn}) and that of the DBI term in \eqref{eq:action}. The result is that the relative sign between the two terms does not change, preserving the form of \eqref{eq:Sd3final}, but the overall sign does, matching the result of \cite{Aharony:2021zkr} on the first branch.} For $T^{1,1}$, \eqref{eq:D3_probe_Mysterious} gives 
\begin{equation}
    S_{\text{D3}}^{T^{1,1}}(\mathcal{D}_{\tau}\times l_I) = \frac{9\pi}{4} N \frac{\varphi^2}{\omega_1\omega_2} \, .
\end{equation}


\subsection{Surface defect contributions}

As a final application of our results, we will consider probing the black hole with a $\tfrac{1}{2}$-BPS surface defect \cite{Chen:2023lzq, Cabo-Bizet:2023ejm}. The inclusion of the defect acts as a non-perturbative order parameter for the large $N$ phase transition between the black hole and thermal AdS$_5$. In this example the brane wraps a non-compact three-cycle inside the black hole and a circle of the internal space. Since the brane extends out to the boundary of AdS$_5$ one should think of it as the insertion of an extended operator in the dual CFT. 

More precisely, the probe brane wraps the cigar and one of the circles of the black hole and another circle in the internal space. In total there are $2\times d$ such four-cycles to consider. 
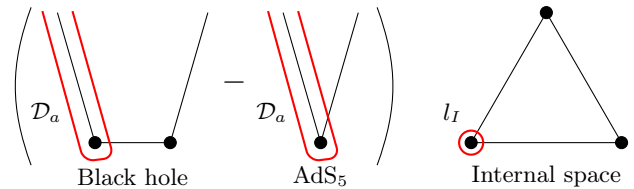
\begin{figure}
    \centering
    \begin{tikzpicture}

\begin{scope}[xshift=6 cm]

\draw[] (-1.85,1.8)  to[out=250,in=110] (-1.85,-0.3);
        
    \draw[] (-1.5,1.73)--(-1,0)--(0,0)--(.5,1.73);

    \filldraw[] (0,0) circle (.08);
    \filldraw[] (-1.,0) circle (.08);

    \draw[rounded corners=4pt, draw=red, thick]  (-1.7,1.75)--(-1.15,-0.25)--(-.75,-0.2)--(-1.3,1.75);

    \node at (-1.65,0.4) {$\cD_a$};
    \node at (-0.5,-0.45) {Black hole};

    \end{scope}

    \begin{scope}[xshift=9 cm]
        
    \draw[] (-1.5,1.73)--(-1,0)--(-0.5,1.73);

    \filldraw[] (-1.,0) circle (.08);

    \draw[rounded corners=4pt, draw=red, thick]  (-1.7,1.75)--(-1.15,-0.25)--(-.75,-0.2)--(-1.3,1.75);

    \node at (-1.65,0.4) {$\cD_a$};
    \node at (-1,-0.45) {$\text{AdS}_5$};

    \node at (-2.15,0.8) {\large{$-$}};

    \draw[] (-0.25,1.8)  to[out=290,in=70] (-0.25,-0.3);

    \end{scope}

    \begin{scope}[xshift=11 cm]
        
    \draw[] (-1,0)--(1,0)--(0,1.73)--(-1,0);

    \filldraw[] (1.,0) circle (.08);
    \filldraw[] (-1.,0) circle (.08);
    \filldraw[] (0,1.73) circle (.08);
    \draw[thick,red] (-1.,0) circle (0.16);

    \node at (-1.25,0.4) {$l_I$};
    \node at (0,-0.45) {Internal space};

    \end{scope}

    \end{tikzpicture}

    \caption{The probe brane wraps a non-compact three-cycle in the black hole. The result is regularized by subtracting the thermal AdS result. }
    \label{fig:BHandAdScigar}
\end{figure}
The cycle is non-compact, so we must regularize the action. In principle one should add counterterms and perform holographic renormalization to obtain a finite result, however following \cite{Chen:2023lzq}, we  regularize by using background subtraction with the corresponding thermal AdS$_5$ result, as depicted in figure \ref{fig:BHandAdScigar}, and following \cite{Colombo:2025ihp,Colombo:2025yqy}. This also removes the constant vacuum contribution associated with the supersymmetric Casimir energy (see \cite{Cassani:2019mms} for a discussion), which therefore does not affect the holographic relations below.

Consider the contribution to the probe brane action from the black hole. There is a single leaf on the black hole to sum over due to the non-compact nature of the cycle. 
To perform the localization we introduce an auxiliary and arbitrary vector field $Y$ which is not collinear to $V$ except at the leaves. 
\footnote{We emphasize that the auxiliary vector field $Y$ is not a choice of regularization, rather it is a requirement for odd-dimensional localization, see \cite{Goertsches:2015vga}, and should drop out after including all possible fixed point and boundary terms.} 
Then the contribution from the black hole is 
\begin{align}
        S_{\text{D3}}(\mathcal{D}_a\times l_I)&=\frac{\mu_{\text{D3}}}{2}\int_{\mathcal{D}_a}\sigma\wedge \dd \sigma \int_{l_I}\eta\\
        &=\frac{4a_{\text{CFT}}}{N}\frac{ (V\hook \sigma)^2(Y,\vBH_a,\vBH_{0})}{(V,\vBH_a,\vBH_{0})(V,Y,\vBH_a)}\int_{l_I}\eta\, ,\nonumber
\end{align}
plus divergent boundary contributions. 
For thermal AdS$_5$, with $V$ still given as in \eqref{eq:BHV}, we find
\begin{align}
        S_{\text{D3}}^{\text{AdS}}(\mathcal{D}_a\times l_I)&=\frac{4a_{\text{CFT}}}{N}\frac{ (V\hook \sigma)^2(Y,\vBH_1,\vBH_{2})}{(V,\vBH_1,\vBH_{2})(V,Y,\vBH_a)}\int_{l_I}\eta\, ,
\end{align}
with the same divergent boundary contributions as the black hole result. 
Subtracting the thermal AdS$_5$ result from the black hole result, we obtain a finite value, independent of the choice of $Y$, for $a=1,2$
\begin{equation}
\label{eq:FiliAndRichie}
S_{\text{D3}}^{\text{reg}}(\mathcal{D}_a\times l_I) = -\ii \frac{a_{\text{CFT}}}{N}\frac{4}{(\xi, v_{I-1},v_{I})} \frac{\varphi^2}{\omega_{a}}\, .
\end{equation}
From the viewpoint of the four-dimensional $\cN=1$ superconformal field theory on $S^1\times S^3$, this expression represents a new prediction for the large $N$ limit of the expectation value of a $\frac{1}{2}$-BPS superconformal defect $\cD$ wrapping the thermal circle and a maximal circle in $S^3$ (ignoring the backreaction on the geometry). The concrete description of the superconformal defect is complicated and generically unknown, but \cite{Chen:2023lzq} conjectured that in the large $N$ limit
\begin{equation}
    \log \langle \cD \rangle \sim 
    \frac{1}{6} b_{\rm 2d}(\cD) \frac{\varphi^2}{\omega_a}\, ,
\end{equation}
where $b_{\rm 2d}(\cD)$ is the defect central charge. Since $\log \langle \cD \rangle = \ii S^{\rm reg}_{\rm D3}(\mathcal{D}_a\times l_I)$, comparing with \eqref{eq:FiliAndRichie}, we find that in the large $N$ limit 
\begin{equation}
    b_{\rm 2d}(\cD) \sim 24 \frac{a_{\text{CFT}}}{N}\frac{1}{(\xi, v_{I-1},v_{I})} \, .
\end{equation}
In particular, this reproduces the behavior $b_{\rm 2d}(\cD) \sim 6N$ of the Gukov--Witten operators in $4d$ SYM discussed in \cite{Chen:2023lzq} (that is, for the case of an $S^5$ internal space), and indeed \eqref{eq:FiliAndRichie} reduces to 
\begin{equation}
  S_{\text{D3}}^{\text{reg}, S^5}(\mathcal{D}_a\times l_I) = - \ii N \frac{\varphi^2}{\omega_a} \, ,
\end{equation}
in perfect agreement with the results in \cite{Chen:2023lzq}. The choice of internal space $T^{1,1}$, instead, would give
\begin{equation}
  S_{\text{D3}}^{\text{reg}, T^{1,1}}(\mathcal{D}_a\times l_I)  =  - \ii N \frac{9}{8} \frac{\varphi^2}{\omega_a} \, .
\end{equation}

\bigskip

\section{Concluding remarks}

In this work we have developed a new approach to computing actions of probe D3-branes in ten-dimensional type IIB backgrounds obtained by uplifting a supersymmetric solution of five-dimensional minimal gauged supergravity on a Sasaki--Einstein five-manifold. By reformulating the $\kappa$-symmetry condition in terms of spinor bilinears, we obtained a simple and universal expression for the probe brane action, given in equation \eqref{eq:Sd3final}. This expression naturally lends itself to evaluation via equivariant localization, allowing us to bypass the need for explicit solutions. This provides a systematic framework for studying non-perturbative objects in holography while also significantly broadening the applicability of localization techniques in supergravity.

We applied this formalism to the supersymmetric Kerr--Newman-AdS$_5$ black hole uplifted on an arbitrary toric SE$_5$, and we computed the action for several families of probe D3-branes. In particular, we obtained general expressions in terms of toric data for branes corresponding to non-perturbative contributions to the dual superconformal index and holographic duals to supersymmetric defects in the dual field theory.  Our results reproduce known expressions in the case of SE$_5 = S^5$, and provide new predictions for a broad class of theories arising from toric Sasaki--Einstein manifolds.

We have only discussed the uplift of a single, physically relevant solution of minimal gauged supergravity. Extending to more general solutions or more general theories, such as the STU model, would allow us to probe even more refined observables. 
Our results suggest that equivariant localization provides a powerful tool for probing non-perturbative aspects of holography in a wide range of settings and we will report on these directions shortly \cite{toappear}.

\begin{acknowledgments}
\noindent We thank Jerome Gauntlett and James Sparks for helpful discussions, and James Sparks for comments on the draft.
AL and CC thank the GGI for hospitality in the final stages of this work.
PBG is supported by the SNSF Ambizione grant PZ00P2\_208666. 
AL is supported by a Palmer Scholarship.
\end{acknowledgments}

\newpage

\bibliography{biblio.bib}

\appendix

\section{Spinor decomposition and bilinears}
\label{app:supplementary_material}

We provide some background definitions for the formula \eqref{eq:Sd3final} for the D3-brane action in terms of spinor bilinears, while leaving a more detailed examination to \cite{toappear}. 
In analogy with the factorization of the ten-dimensional metric \eqref{eq:10d_Split_Metric}, the ten-dimensional Majorana--Weyl Killing spinors decompose as
\begin{equation}
\label{eq:10dspinors}
\begin{split}
    \varepsilon_1 &= \frac{1}{2}(\epsilon\otimes \chi\otimes \Theta+\text{M.c.})\, ,\\ 
    \varepsilon_2 &= \frac{1}{2\ii}(\epsilon\otimes \chi\otimes \Theta-\text{M.c.})\, ,
\end{split}
\end{equation}
where $\epsilon$ is a spinor on $M_5$ that becomes the Killing spinor of the five-dimensional gauged supergravity, $\chi$ is a Killing spinor on the Sasaki--Einstein manifold and $\Theta=(1,0)$ is a two-component auxiliary spinor necessary for the decomposition. We denote the Majorana conjugate by $\text{M.c.}$ which makes the two spinors manifestly Majorana. We take the ten-dimensional Clifford matrices to be
\begin{equation}
   \Gamma_{M}= \begin{cases}
        \gamma_m\otimes 1\otimes \sigma_1 & M=0,1,2,3,4,\\
        1\otimes \rho_{\mu}\otimes \sigma_2& M=5,6,7,8,9,
    \end{cases}
\end{equation}
where the $\{\gamma_m\}$ are a basis of the five-dimensional Lorentzian Clifford algebra, $\{\rho_\mu\}$ are a basis for the five-dimensional Euclidean Clifford algebra and $\sigma_A$ are the usual Pauli matrices.

One can now form spinor bilinears from both sets of the five-dimensional spinors. From the Killing spinor of five-dimensional gauged supergravity one constructs the spinor bilinears
\begin{equation}
    f\equiv\ii\, \overline{\epsilon}\epsilon\, ,\quad V^{\flat}\equiv \overline{\epsilon}\gamma_{(1)}\epsilon\, ,\quad X\equiv-\overline{\epsilon}\gamma_{(2)}\epsilon\, ,
\end{equation}
where $\gamma_{(p)}$ denotes contraction with $p$ vielbein. Similarly, on the Sasaki--Einstein manifold one can construct analogous spinor bilinears as
\begin{equation}
    \overline{\chi}\chi=1 \, ,\quad \eta=\xi^{\flat}\equiv\overline{\chi}\rho_{(1)}\chi \, ,\quad J=-\ii \overline{\chi}\rho_{(2)}\chi\, .
\end{equation}
One can now plug the decomposition of the ten-dimensional spinors in \eqref{eq:10dspinors} into \eqref{eq:Kdef} and \eqref{eq:Phidef} and rewrite the result in terms of the above five-dimensional spinor bilinears. This leads to \eqref{eq:Kin5} and \eqref{eq:Phi3in5}, respectively.

\end{document}